\documentclass{article}
\usepackage{epsfig}
\usepackage{amssymb}
\newcommand{\bfr}{\begin{flushright}}
\newcommand{\efr}{\end{flushright}}
 
\begin{document}
\title{Effect of Self-Interaction on Vacuum Energy for Yang-Mills 
System in Kaluza-Klein Theory
}
\author{Kiyoshi Shiraishi\\
Akita Junior College, Shimokitade-sakura, Akita-shi,\\ Akita 010,
Japan\\ 
and\\
Satoru Hirenzaki\\
The Institute of Physical and Chemical Research,\\ 
Wakoh-shi, Saitama 35141, Japan
}
\date{CHINESE JOURNAL OF PHYSICS VOL. 30, NO. 4, pp. 431--436\\ AUGUST
1992 }
\maketitle
\begin{abstract}
We calculate the vacuum energy for Yang--Mills (YM) system in the
background space-time $M^4 \times S^3$, taking the effect of
self-interaction of the YM fields into account. The compactification
scale obtained by Candelas--Weinberg mechanism becomes large if the YM
coupling is large. The case with an extra space $S^3/Z_2$ is also
considered, and it is shown that the vacuum associated with broken
gauge symmetry is unstable.
\end{abstract}

In Kaluza-Klein theories,\cite{1} the effect of vacuum polarizations is
expected to play important roles in stabilizing the compact
space\cite{2} as well as in determining the gauge symmetry in four
dimensions.\cite{3}

However, in almost all the calculations of quantum effects in the Kaluza-Klein scheme (except
for Ref.~\cite{4}, only the one-loop approximation is taken into
consideration. Thus the calculation carried out so far might not
reflect the quantum nature of self-interacting system such as
gravitation and YM fields which are inevitably contained in string
theory. 

The difficulty in going beyond one-loop is caused partly from the lack of appropriate
method in computing the higher-loop contribution unambiguously in arbitrary dimensions.
In the previous paper,\cite{6} the present authors have developed a
technique to handle the YM interaction in vacuum diagrams when the
background space-time metric has an extra space $S^1$.

It is found that the approximation used in the calculation is more suitable for higher dimensions.
We can study the higher-loop effects by our method similarly to the one-loop technique, even
if the extra space has a complicated structure.
In the present paper, we show the vacuum energy for $SU(3)$ YM system
defined in the background space-time $M^4 \times S^3$, where $M^4$ is
the flat Minkowski space-time and $S^3$ is the three sphere which has
non-vanishing curvature.

We fast review our approximation scheme described in Ref.~\cite{6}. One
can find that the contributions of the vacuum graphs including only YM
four-point interaction dominate over the ones of the other graphs of
the same order in $g^2$ (where $g$ is the YM coupling constant) for
large dimensionality, $D$. It arises because the trace of the metric at
a closed loop yields a factor $D$ for a vacuum diagram. The graphs which
include three-point interactions with derivative couplings are
sub-dominant. Therefore it is conceivable that for large $D$ only
four-point interaction is important and this simplifies the treatment
of YM interactions of higher order. As a check, we have shown in
Ref.~\cite{6} that the result of the one- and two-loop order of the
vacuum energy for YM system in $S^2 \times R^d$ can be reconstructed
when
$g^2\ll 1$ by using our method. In our approximation scheme, we start
with the YM action where three-point interactions are omitted.

Owing to this simplification, the calculation of the vacuum energy becomes very
transparent. We can utilize auxiliary fields \cite{7} to treat the
non-linear interaction in the reduced Lagrangian.

A difference from the previous work is the choice of the gauge group. For general gauge
group, a quartic interaction which we preserve in the reduced action is
\begin{equation}
\sum_{a, b, c}f^{abc}f^{abc}A^b_\mu A^{b\mu}A^c_\nu A^{c\nu}\,,
\label{1}
\end{equation}
where $f^{abc}$ is the structure constant of the gauge group. The other
types of interactions are discarded because they cannot produce the
graphs of leading contribution (see Ref.~\cite{6}). Then we can give
the expression of the effective action, in which the gauge fields
appear in bilinear form, using auxiliary fields.

Now let us concentrate on the specific case, $SU(3)$ YM theory. Here, we
explicitly show the structure of the auxihary fields.

For concreteness, we use the Gell-Mann's notation.\cite{8} We find
\begin{equation}
\sum_a f^{abc}f^{abc}=\left(
\begin{array}{cccccccc}
0 & 1 & 1 & 1/4 & 1/4 & 1/4 & 1/4 & 0\\
1 & 0 & 1 & 1/4 & 1/4 & 1/4 & 1/4 & 0\\
1 & 1 & 0 & 1/4 & 1/4 & 1/4 & 1/4 & 0\\
1/4 &1/4 &1/4 &0 &1 &1/4 &1/4 &3/4\\
1/4 &1/4 &1/4 &1 &0 &1/4 &1/4 &3/4\\
1/4 &1/4 &1/4 &1/4 &1/4 &0 &1 &3/4\\
1/4 &1/4 &1/4 &1/4 &1/4 &1 &0 &3/4\\
0 &0& 0& 3/4& 3/4& 3/4& 3/4& 0
\end{array}
\right)\,,
\label{2}
\end{equation}
where the column and row correspond to the suffices $b$ and $c$. The
inverse of this matrix is
\begin{equation}
M_{ab}\equiv\left(
\begin{array}{cccccccc}
-1/2 & 1/2 & 1/2 & 0 & 0 & 0 & 0 & -1/6 \\
1/2 & -1/2 & 1/2 & 0 & 0 & 0 & 0 & -1/6 \\
1/2 & 1/2 & -1/2 & 0 & 0 & 0 & 0 & -1/6 \\
0 & 0 & 0 & 0 & 1 & -1/2 & -1/2 & 1/3 \\
0 & 0 & 0 & 1 & 0 & -1/2 & -1/2 & 1/3 \\
0 & 0 & 0  & -1/2 & -1/2 & 0 & 1& 1/3 \\
0 & 0 & 0  & -1/2 & -1/2 & 1 & 0& 1/3 \\
-1/6 & -1/6 & -1/6 & 1/3 & 1/3 & 1/3 & 1/3& -1/2
\end{array}
\right)\,,
\label{3}
\end{equation}
and then we use auxiliary fields $\chi^a (a=1,\dots,8)$ to get the
bilinear form with respect to the YM field. We rewrite the Lagrangian as
\begin{equation}
\frac{1}{2}\sum_{\mu\nu}\sum_a(D^B_\mu a^a_\nu)^2+
\frac{1}{2}\sum_{\mu}\sum_a \chi^a a^a_\mu a^a_\mu-\frac{1}{4g^2}
\sum_{a b}\chi^a M_{ab}\chi^b\,,
\label{4}
\end{equation}
where $a^a_\mu$ is the quantum fluctuation of the gauge field while
$D^B_\mu$ means the covariant derivative involving the background
classical fields. 

Now in order to get the vacuum energy, it is
sufficient to find the effective potential $V(\chi)$ in the one-loop
approximation within the auxiliary fields. We have
\begin{equation}
V(\chi)=-\frac{1}{8g^2}\chi^a M_{ab}\chi^b-\frac{i}{2(Vol.)}
\ln \det(-D^B_\mu D^{B\mu}+\chi^2)\,,
\label{5}
\end{equation}
where $(Vol.)$ stands for the space-time volume, $\int d^4x$.

In the expression (\ref{5}), we implicitly take the formal determinant
as the one which contains not only transverse and longitudinal
components of the vector field but also Faddeev-Popov ghost fields.

The final expression should be obtained from $V(\chi)$ by eliminating
the auxiliary field x with the help of the equation of motion.

The regularization and the calculation technique of the one-loop vacuum energy is the
same as Ref.~\cite{2,9,10,11}. Particularly, the calculation for
vector fields is parallel to Ref.~\cite{10,11}. 

First we calculate the
vacuum energy for $SU(3)$ YM fields in $M^4 \times S^3$, where no
background gauge field exists. The symmetry suggests that the auxiliary
fields should be set to an equal value, $\chi^1=\chi^2=\dots=\chi$.
Thus we have to solve one equation for the auxiliary field to obtain the
value of the vacuum energy.

\begin{figure}[ht]
\begin{center}
\includegraphics[width=6cm]{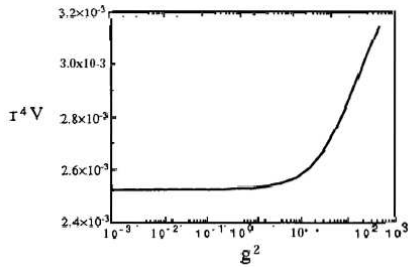}
\caption{The vacuum energy for the YM system in $M^4 \times S^3$ is
plotted against the dimensionless parameter $\tilde{g}^2=g^2/r^3$.
}
\label{f1}
\end{center}
\end{figure}

The result for the vacuum energy normalized by $r^4$ (where $r$ is the
radius of $S^3$) is given in Fig.~1. The equation of motion for the
auxiliary fieldx has been solved by numerical calculation. 

In Fig.~1,
the vacuum energy slightly increases as a dimensionless combination
$\tilde{g}^2=g^2/r^3$ grows. Obviously, the limit $g\rightarrow 0$
yields the free case, i.e., the vacuum energy for eight species of
abelian gauge fields (cf. Ref.~\cite{10}).

Next let us suppose the vacuum energy in $M^4 \times S^3/Z_2$. In this
case, allowed Kaluza-Klein
modes of the YM field are restricted and dependent on the background
classical gauge configurations in the extra space.\cite{10}

We consider the following two possible vacuum configurations of gauge
fields as the following; one is the trivial case, $\langle
A_m\rangle=0$ and the other is $\langle
A_\phi\rangle={\rm diag}. (1, 1, -2)$, where  
expresses the direction of the one of the azimuthal angle of $S^3$ (see
Ref.~\cite{11}). In the vacuum associated with the latter configuration,
the four-dimensional gauge symmetry is reduced to $SU(2) \times U(1)$.

Because of the symmetry, we can set $\chi^1=\chi^2=\chi^3=\chi^8\equiv
\chi^A$ and
$\chi^4=\chi^5=\chi^6=\chi^7=\chi^B$ then the quadratic part of the
auxiliary fields in the vacuum energy is given as 
\begin{equation}
\frac{1}{8g^2}(\chi^A\chi^B)\,.
\label{6}
\end{equation}

We have to solve two
equations of the auxiliary fields simultaneously for each vacuum
configuration.

\begin{figure}[ht]
\begin{center}
\includegraphics[width=6cm]{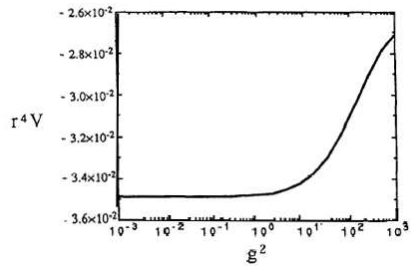}
\caption{The vacuum energy of the trivial vacuum for the YM system in
$M^4\times S^3/Z_2$, is plotted against the dimensionless parameter
$\tilde{g}^2=g^2/r^3$.}
\label{f2}
\end{center}
\end{figure}

For the trivial vacuum, where we can set $\chi^A=\chi^B$, the results
of the numerical calculation for the vacuum energy normalized by $r^4$
is shown by Fig.~2, similarly to the case of $S^3$. In the non-trivial
case, however, we cannot find the solution of the equation of motion of
auxiliary fields for positive $\chi^A$ and $\chi^B$. This indicates the
quantum instability of the non-trivial vacuum, because $\chi$'s stand
for ``effective squared-masses''  of the gauge bosons in physical
meaning. (see Eq.~(\ref{4}).)

To summarize, we have computed vacuum energies for $SU(3)$ Yang-Mills
theory in partially compactified background geometry in seven
dimensions.

The details in the calculation of the vacuum energy will be reported
elsewhere.

 The vacuum energy for the finite coupling constant in the background
$M^4\times S^3$ is shown in Fig~1. The compactification scheme which
utilizes the positive vacuum energy and cosmological constant\cite{2}
may not be drastically changed by the self-interaction effect. Of
course, the condition for the tuning of the cosmological constant to
cancel the four-dimensional vacuum energy would be slightly modified.
If $\tilde{g}=10$ the compactification scale is expected to become about
several to ten percent larger than the free-fleld case.

As for the Hosotani mechanism in the space-time $M^4 \times S^3/Z_2$, it
is found that the interaction forces the non-trivial vacuum to be
unstable. The self-interaction plays the crucial role in the Hosotani
mechanism.

Unfortunately, we cannot advocate the quantum instability of non-trivial vacua for general
gauge groups and manifolds. We hope to study the stability of the vacuum in this sense in general
self-interacting fields in various dimensions and will report the result elsewhere.

\section*{ACKNOWLEDGMENT}
The authors would like to thank A. Nakamula for some useful comments.
KS is indebted to Iwanami F\=ujukai for financial support.


\end{document}